\documentclass[twocolumn]{aastex63}

\usepackage{grffile}
\usepackage{graphicx}
\usepackage{amsmath}
\usepackage{gensymb}
\received{XX}
\revised{\today}
\accepted{XX} 
\submitjournal{ApJ}

\shorttitle{Sample article}
\shortauthors{Li et al.}

\newcommand{\oiii}{[\ion{O}{3}]$_{\rm 88 \mu m}$}

\newcommand{\lsun}{$L_{\sun}$}

\newcommand{\cotwo}{CO$(2-1)$}
\newcommand{\coone}{CO$(1-0)$}
\newcommand{\ci}{[\ion{C}{1}]$_{\rm 369 \mu m} $}
\newcommand{\cii}{[\ion{C}{2}]$_{\rm 158 \mu m} $}
\newcommand{\cosix}{CO$(6-5) $}
\newcommand{\cosev}{CO$(7-6) $}
\newcommand{\qso}{J0305$-$3150}
\newcommand{\kmps}{$\rm km\ s^{-1}$}
\newcommand{\mjypb}{$\rm mJy \ beam^{-1}$}
\newcommand{\ujypb}{$\rm \mu Jy \ beam^{-1}$}
\newcommand{\jykmps}{$\rm Jy\ km\ s^{-1} $}
\newcommand{\msun}{$\rm M_{\sun}$}

\begin{document}

\title{Spatially resolved molecular interstellar medium in a $z=6.6$ quasar host galaxy}

\correspondingauthor{Jianan Li }
\email{jiananl@pku.edu.cn }
\author[0000-0002-1815-4839 ]{Jianan Li}
\affiliation{Kavli Institute for Astronomy and Astrophysics, Peking University, Beijing 100871, China}
\affiliation{Department of Astronomy, Tsinghua University, Beijing 100084, China}

\author[0000-0001-9024-8322]{Bram P. Venemans}
\affiliation{Max-Planck Institute for Astronomy, K{\"o}nigstuhl 17, D-69117 Heidelberg, Germany}
\affiliation{Leiden Observatory, Leiden University, PO Box 9513, 2300 RA Leiden, The Netherlands}

\author[0000-0003-4793-7880]{Fabian Walter}
\affil{Max-Planck Institute for Astronomy, K{\"o}nigstuhl 17, D-69117 Heidelberg, Germany}
\affil{National Radio Astronomy Observatory, Pete V. Domenici Array Science Center, P.O. Box O, Socorro, NM 87801, USA}

\author[0000-0002-2662-8803]{Roberto Decarli}
\affiliation{INAF -- Osservatorio di Astrofisica e Scienza dello Spazio, via Gobetti 93/3, 40129 Bologna, Italy}

\author[0000-0003-4956-5742]{Ran Wang}
\affiliation{Kavli Institute for Astronomy and Astrophysics, Peking University, Beijing 100871, China}

\author[0000-0001-8467-6478]{Zheng Cai}
\affiliation{Department of Astronomy, Tsinghua University, Beijing 100084, China}



\begin{abstract}
We present high spatial resolution ($\sim$0\farcs4, 2.2kpc) observations of the \cosix{}, \cosev{} and  \ci{} lines and dust continuum emission from the interstellar medium in the host galaxy of the quasar \qso{} at $z=6.6$. 
These, together with archival \cii{} data at comparable spatial resolution, enable studies of the spatial distribution and kinematics between the ISM in different phases. 
When comparing the radial profiles of CO, \cii{} and the dust continuum, we find that the CO and dust continuum exhibit similar spatial distributions, both of which are less extended than the \cii{}, indicating that the CO and dust continuum are tracing the same gas component, while the [CII]158um is tracing a more extended one.
In addition, we derive the radial profiles of the \cii{}/CO, \cii{}/far-infrared (FIR), CO/FIR, and dust continuum $S_{98.7 \rm GHz}/S_{258.1 \rm GHz}$ ratios. 
We find a decreasing  $S_{98.7 \rm GHz}/S_{258.1 \rm GHz}$ ratio with radius, possibly indicating a decrease of dust optical depth with increasing radius.
We also detect some of the ISM lines and continuum emission in the companion galaxies previously discovered in the field around \qso{}.  Through comparing the line-to-line and line-to-FIR ratios, we find no significant differences between the quasar and its companion galaxies.

\end{abstract}


\section{Introduction} \label{sec:intro}

In the past two decades, the Atacama Large Millimeter/submillimeter Array (ALMA), the NOrthern Extended Millimeter Array (NOEMA) and the Karl G. Jansky Very Large Array (JVLA) have revealed detections of the (sub)millimeter dust and multi-phase gas emission in quasar host galaxies in the early universe. Bright CO and \cii{} emission lines are now frequently detected in the host galaxies of the quasars at the highest redshift. The majority of these observations are executed at $\gtrsim 0\farcs 5$ spatial resolution, which trace the global interstellar medium (ISM) properties of these $z\gtrsim 6$ quasars (e.g., \citealt{wang13,wang16}; \citealt{decarli18}; \citealt{yang19}; \citealt{li20a, li20b}). Spatially resolved ISM observations of $z\gtrsim 6$ quasars,  are only available for the brightest ISM emission lines, e.g., the \cii{} line. These already reveal variations of gas kinematics in the $z\gtrsim 6$ quasars, i.e., some of them suggest ordered rotation, others show complex gas kinematics with no clear velocity gradient (e.g., \citealt{shao17}; \citealt{neelman19}; \citealt{venemans19,  venemans20}; \citealt{wang19a}; \citealt{novak20}; \citealt{neeleman21}).
Recently, spatially resolved CO observations at a spatial resolution of $\sim$ 0\farcs2 for a $z= 6.327$ quasar have been obtained by \citet{wang19b}, where they  found a more concentrated spatial distribution of CO compared to the \cii{} line. 

Taking advantage of the high sensitivity of ALMA, a number of companion galaxies have recently been discovered in the field of quasars at $z\gtrsim 6$ (e.g., \citealt{decarli17, decarli18};  \citealt{walter18};  \citealt{mazzucchelli19};  \citealt{neelman19, neelman21};  \citealt{venemans18, venemans20}). These companion galaxies are detected within $\lesssim$ 60 kpc and within $\sim \rm 1000\ km\ s^{-1}$ of the quasar redshift and are often found to be bright in the far-infrared (FIR; 42.5$\sim$122.5$\mu m$) continuum and the \cii{} line. 
The brightest ones have \cii{} luminosities comparable to or even brighter than that of the quasar host galaxies, while the less luminous companions are over an order of magnitude fainter. 
Direct comparisons of the quasar with their companion galaxies provide a unique view on the potential impact of the Active Galactic Nuclei (AGN) on the ISM properties. 
Observations of fine-structure lines and molecular CO suggest similar \oiii{}/FIR and \cii{}/FIR ratios but different CO excitation  between the quasars and their companions (e.g., \citealt{walter18}; \citealt{neelman19}; \citealt{pensabene21}).

The quasar VIKING J030516.92–315056.0 (hereafter \qso{}) is among the FIR brightest quasars at  $z> 6$ with a FIR luminosity of (1.60 $\pm$ 0.06)$\times 10^{13}$ \lsun{} \citep{venemans19}. 
It was also detected in the \cii{} line with a luminosity of (5.9 $\pm$ 0.4)$\times 10^{9}$ \lsun{} \citep{venemans19}.
In ALMA Cycle 2, \citet{venemans17a} detected the \cosix{} and \cosev{} lines.  ALMA observations of the \cii{} line with extremely high spatial resolution ($0.076''$,  410pc) was reported in \citet{venemans19}, which reveal complex gas spatial distribution and kinematics. Two cavities found in the zero velocity channel map as well as the intensity map suggest that the quasar is likely to affect the spatial distribution and kinematics of its surrounding ISM in the host galaxy. 
In addition, three companion galaxies within 40 kpc from the quasar \qso{} were detected in \cii.

In this paper, we present spatially resolved ($\sim$ 0\farcs4, 2.2kpc) ALMA observations of the \cosix{}, \cosev{}, and \ci{} emission lines as well as the dust continuum emission of the quasar \qso. 
These observations probe the molecular ISM, and enable a direct comparison of the spatial distribution and kinematics the ISM emission in different phases, when combined with previous \cii{} observations at similar spatial resolution.  In addition, comparisons of the quasar emission with that of the companion galaxies will enable a study of the impact of the AGN on the gas properties. 
We adopt a standard $\Lambda$CDM cosmology with with $H_{0}=70\ \rm km \ s^{-1}\ Mpc^{-1}$ and $\Omega_{\rm m}=0.3$, throughout this paper.

\section{Observations}
We obtained ALMA observations of the \cosix{}, \cosev{}, and \ci{} emission lines as well as the underlying continuum of \qso{} during 2017 December 04--17 (Cycle 5 program ID 2017.1.01532.S).
45-47 antennas were used in the C43--6 configuration and the baseline length was between 15 and 2517 meters. The total observing time  was 3.67 hours on source.
J2357--5311 was used for flux and bandpass calibration, and the phase calibrator was J0326--3243.
The ALMA Band 3 receiver covered the \cosix{} line in the lower sideband and the \cosev{} and \ci{} lines in the upper sideband, while the remaining two spectral windows were observing the continuum emission.

To study the spatial distribution and kinematics of the ISM in different phases, we also use earlier ALMA data of the \cii{} line, which traces the neutral ISM (Cycle 3 program ID 2015.1.00399.S). This \cii{} data of \qso{} has previously been published in \citet{venemans20}.

All the data were reduced following the standard pipeline and imaged using the TCLEAN task in CASA. 
To obtain a comparable beam size of the \cii{} with that of the CO lines, we employed natural weighting for the \cii{} data and Briggs weighting with a robust parameter of 0.5 for the CO lines in imaging.
This leads to synthesized beam sizes of 0\farcs20 $\times$ 0\farcs20 for \cii{}, 0\farcs44 $\times$ 0\farcs30 for \cosix{} and 0\farcs37 $\times$ 0\farcs26 for \cosev{} and \ci{} in FWHM. 
To match the beam sizes of CO and \cii{}, we downgraded the spatial resolution of the \cii{} data to that of the \cosix{} line using the convolve2D function in CASA.
We used all line-free channels to image the continuum. 
A first order polynomial continuum was subtracted from the datacube using UVCONTSUB task in CASA for spectral line imaging.  We binned the \cii{} line  to 35 \kmps{} width and the resulting rms was  0.28 \mjypb{} per binned channel.
The \cosix{}, \cosev{}, and \ci{} lines were also binned to 35 \kmps{} width and the rms was  0.13 \mjypb{} per binned channel. All line-free channels were used for the continuum imaging. The continuum sensitivities were 5.7 and 20.3\ujypb{} at 98.7 and 258.1 GHz, respectively. 
\section{Results}
\subsection{The quasar}
\begin{figure*}
\centering
\includegraphics[width=1.0\textwidth]{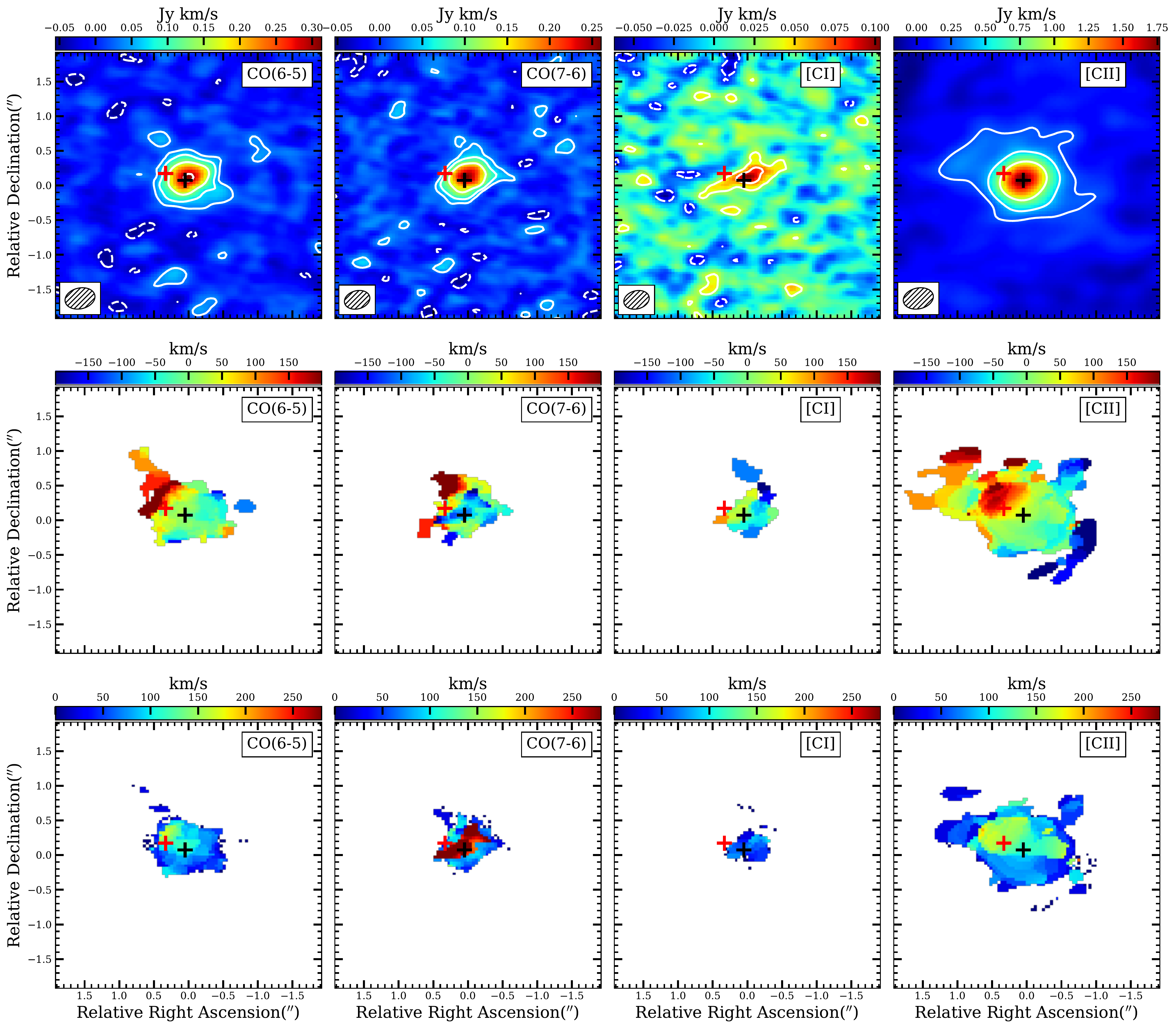}
\caption{Line intensity, velocity, and velocity dispersion maps (from top to bottom) of the quasar \qso. 
The white contours are [--2, 2, 4, 8, 16]$\times \sigma$, $\sigma=0.018$ \jykmps{} for the \cosix{} line, [--2, 2, 4, 8, 16]$\times \sigma$, $\sigma=0.017$ \jykmps{} for the \cosev{} line, [--2, 2, 4]$\times \sigma$, $\sigma=0.020$ \jykmps{} for the \ci{} line, and [--2, 2, 4, 8]$\times \sigma$, $\sigma=0.051$ \jykmps{} for the \cii{} line.
The while ellipse on the lower left represent the size of the beam. The beam sizes from left to right are 0\farcs44$\times$0\farcs30, PA=--77.00 deg, 0\farcs37$\times$0\farcs26, PA=--74.92 deg, 0\farcs37$\times$0\farcs26, PA=--74.90 deg, and 0\farcs44$\times$0\farcs30, PA=--77.00 deg. 
For the velocity and velocity dispersion maps, all the pixels with S/N $>$ 2 are included. The black and red crosses represent the \cii{} peak position for the quasar and C1, respectively.
}
\label{fig1}
\end{figure*}

\begin{figure*}
\includegraphics[width=0.5\textwidth]{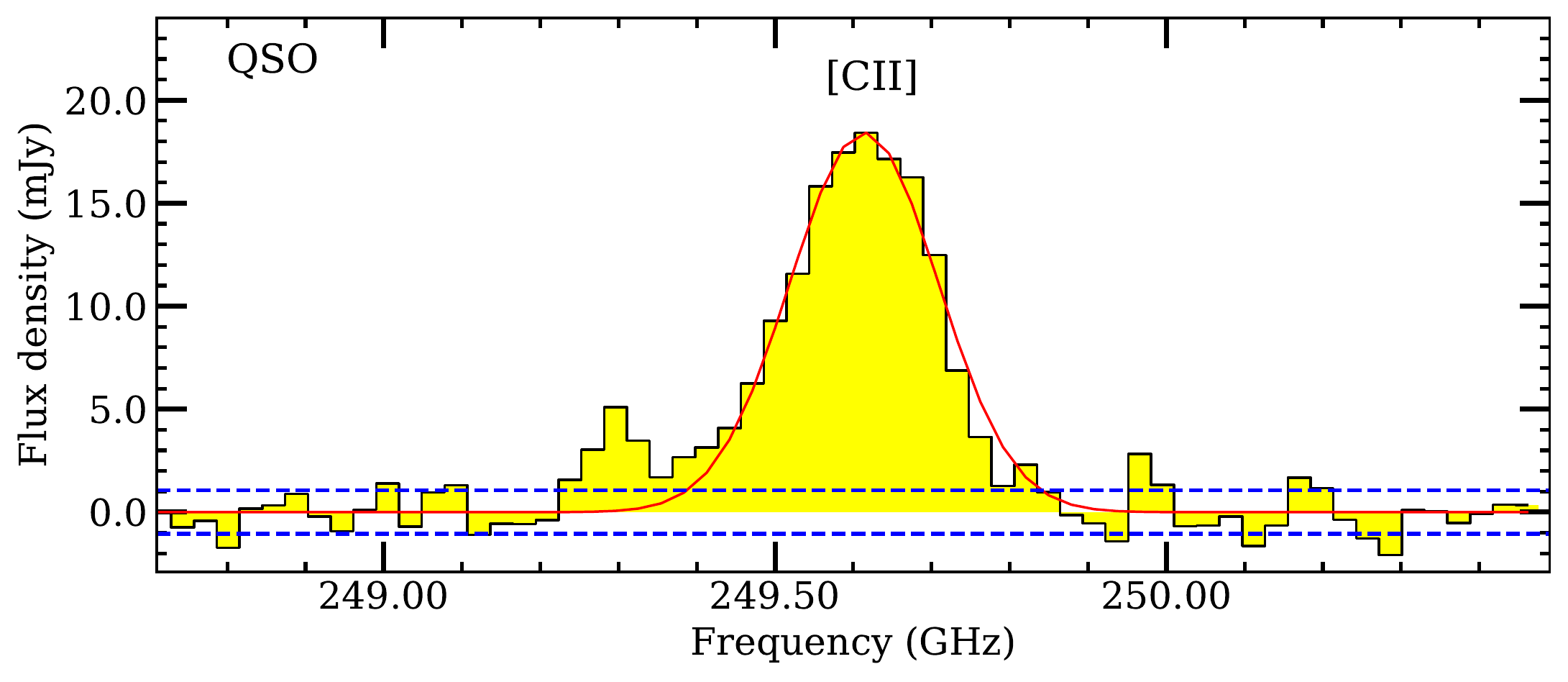}
\includegraphics[width=0.5\textwidth]{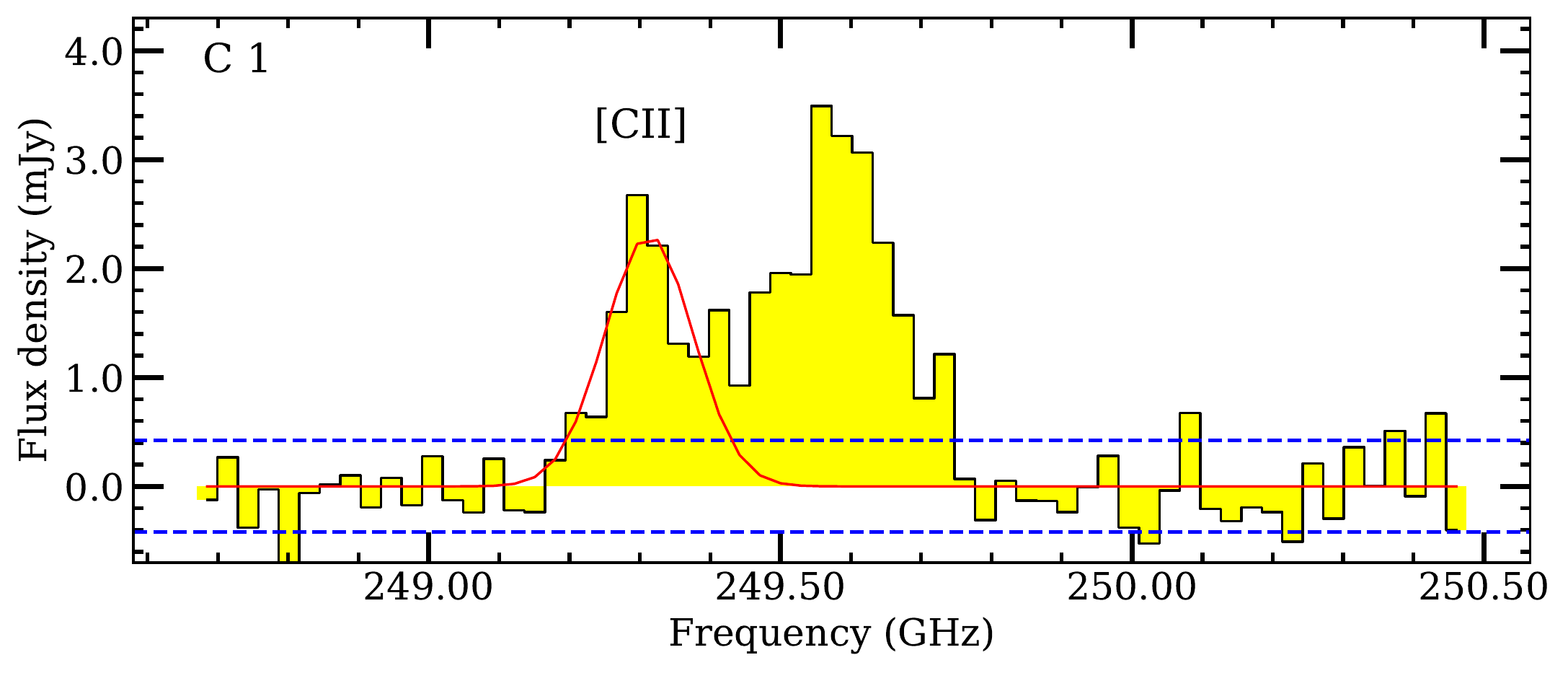}
\includegraphics[width=0.5\textwidth]{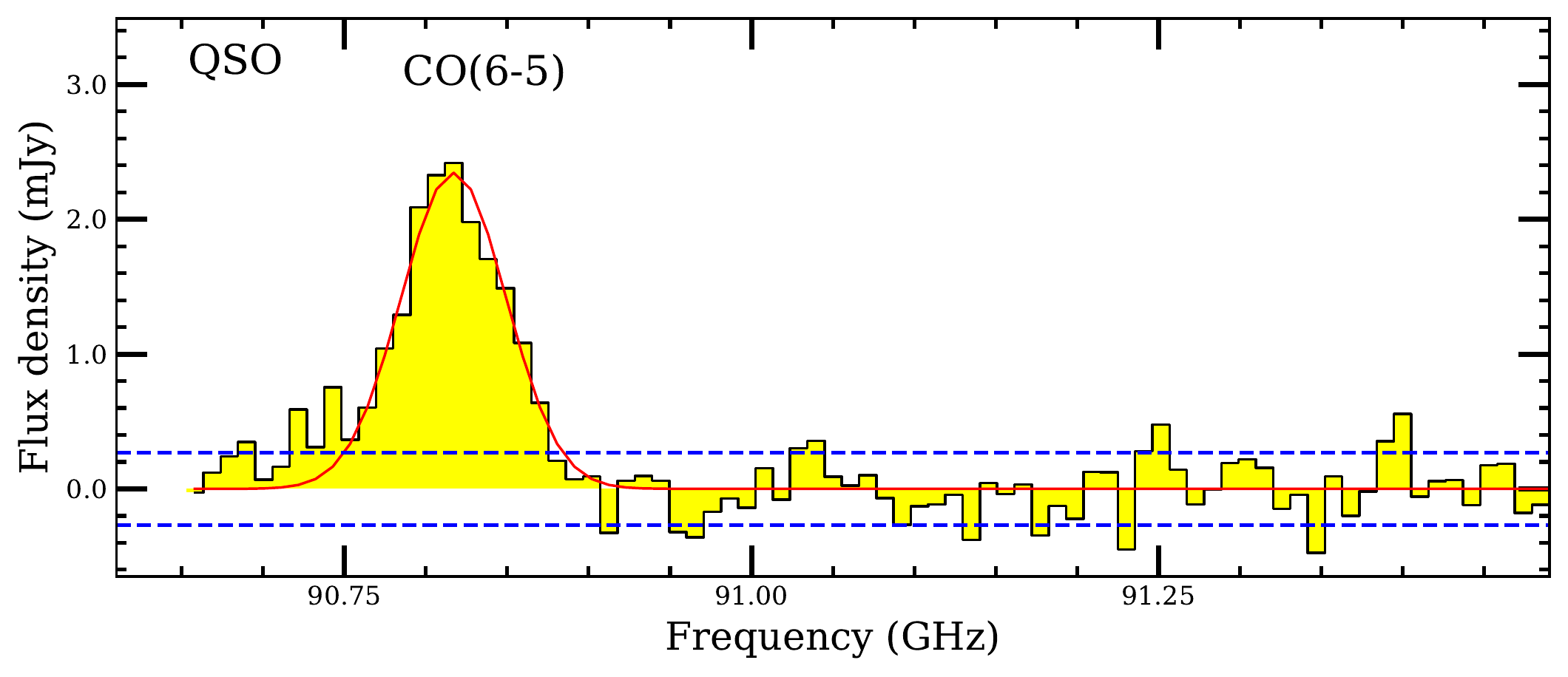}
\includegraphics[width=0.5\textwidth]{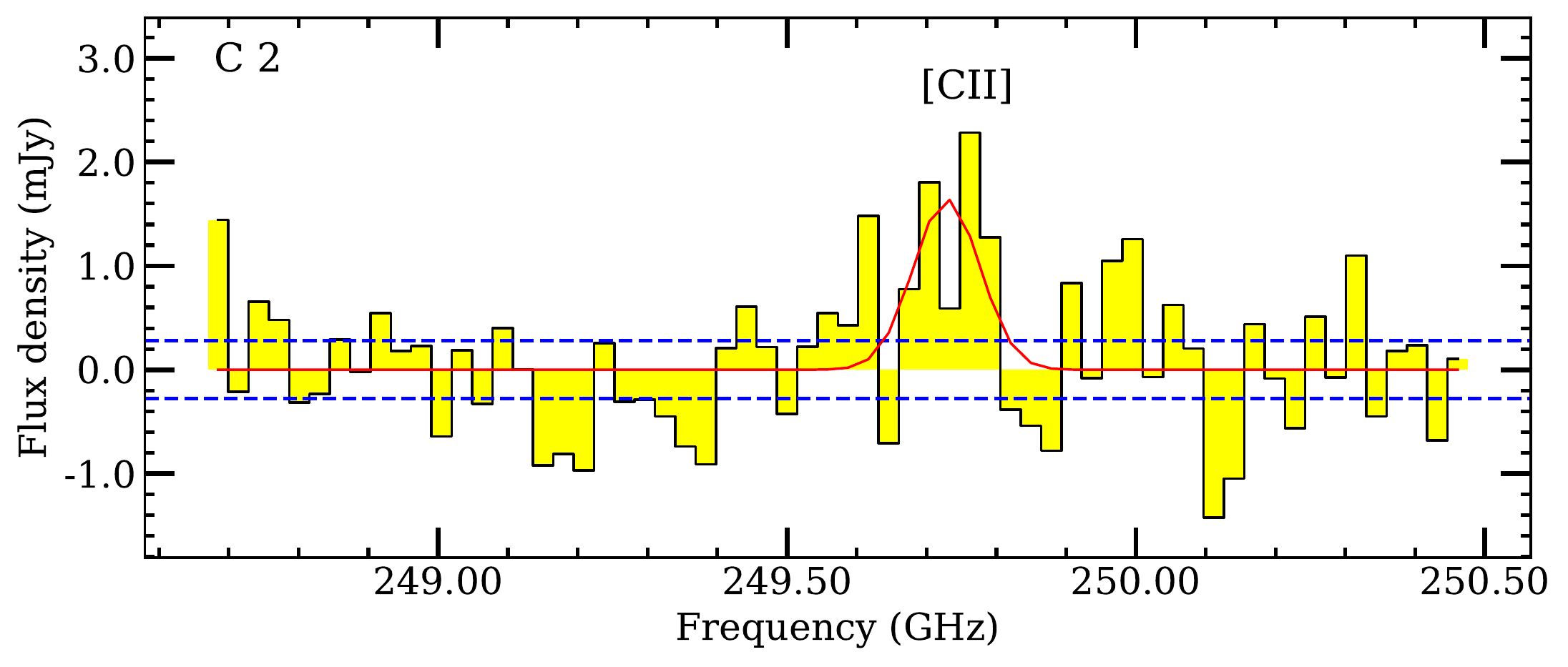}
\includegraphics[width=0.5\textwidth]{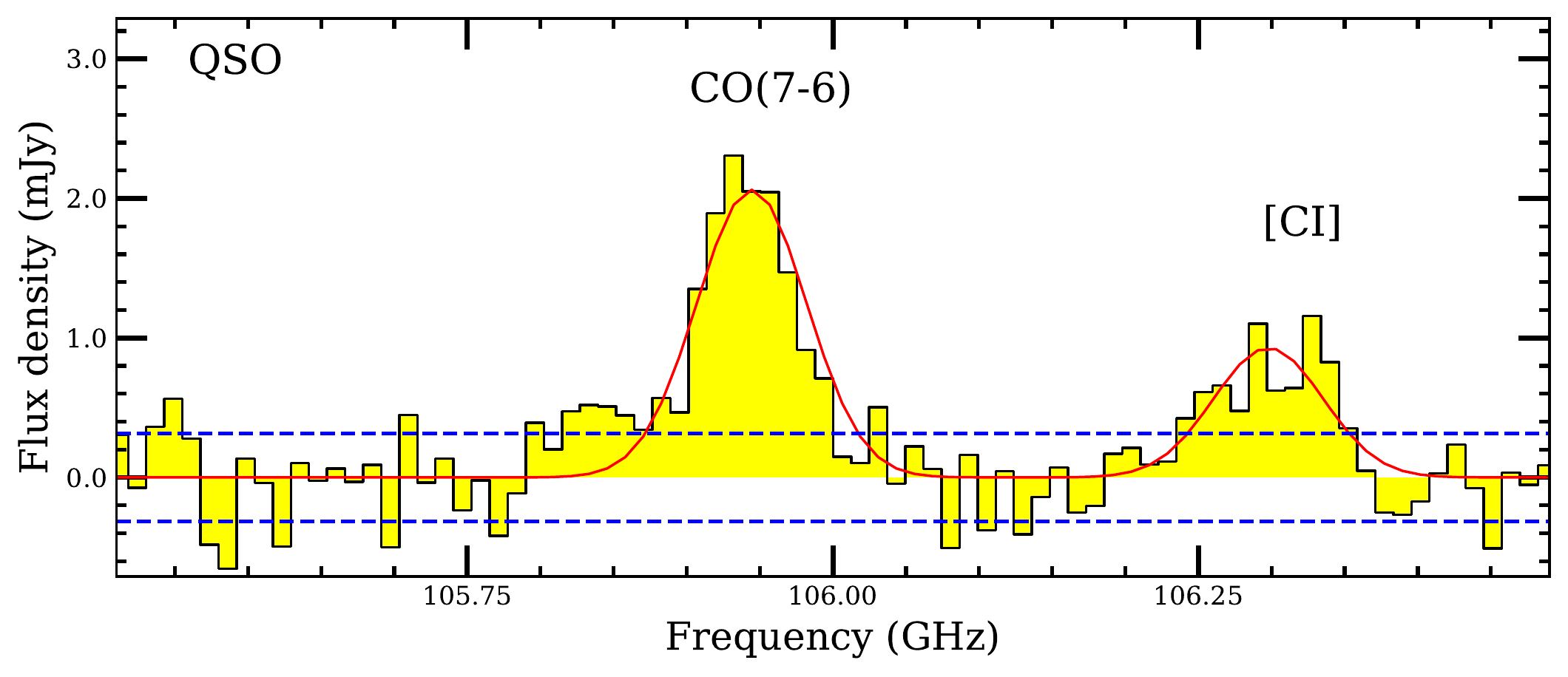}
\includegraphics[width=0.5\textwidth]{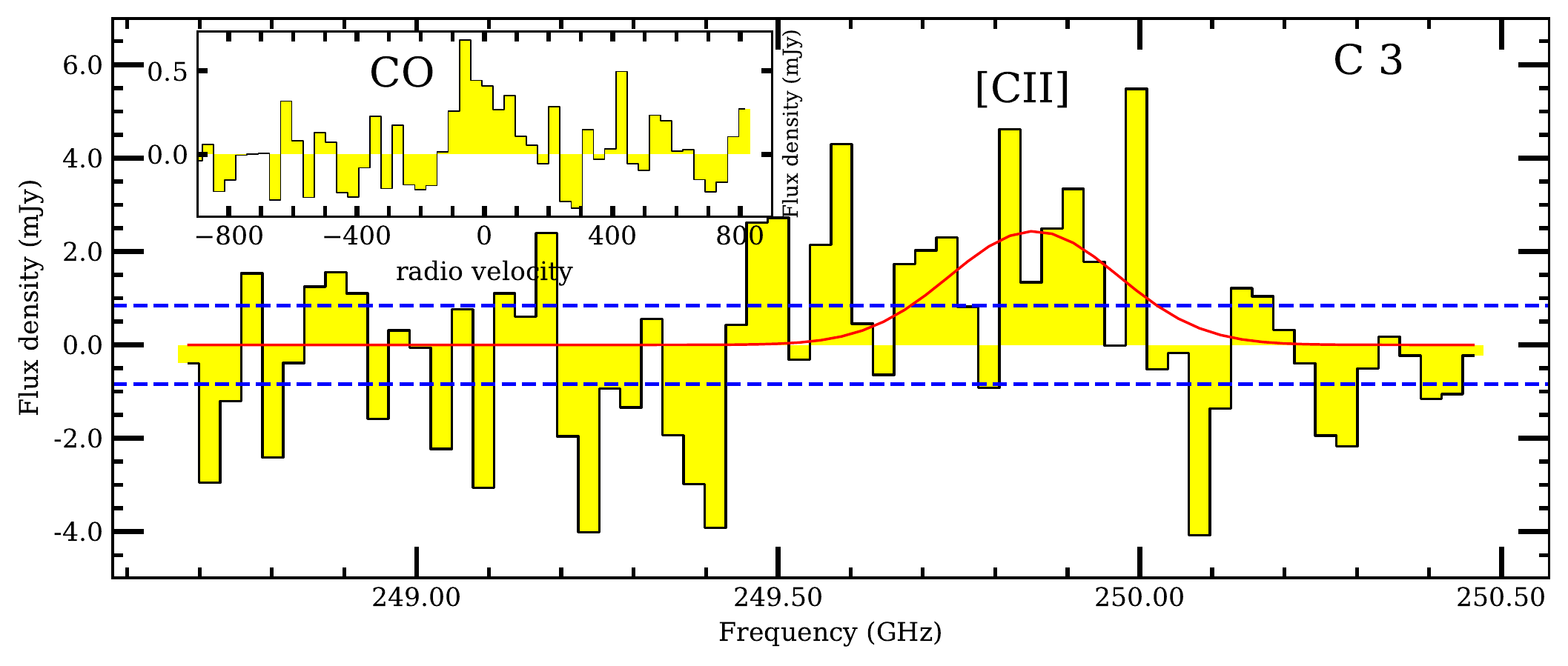}
\caption{Left column: Spectra of \cii,  \cosix, \cosev{} for the quasar \qso. Data are shown in yellow histograms. The red solid lines are Gaussian profile fits to the spectra with the line centers fixed to the \cii{} redshift of $z=6.61391$ \citep{venemans19}. The continuum has been subtracted for each of the spectra here displayed. Right column: \cii{} spectra for the companion galaxies C1, C2 and C3.  Data are shown in yellow histograms. The red solid lines are Gaussian profile fits to the spectra with the line centers fixed to the \cii{} redshift obtained from  \citet{venemans19}.\label{spectra3} The blue dashed lines indicate the noise at $\pm$ 1 $\sigma$ level.
The \cii{} spectrum of C3 is extracted from an aperture with a radius of 0\farcs6.
The spectral channel widths for the quasar and the companion galaxies are 35 \kmps. 
For the source C3,  we present a stacked spectrum of the \cosev{} and \cosix{} lines extracted from the peak positions on the top left of the  \cii{} spectrum panel.}
\end{figure*}

We detect the \cosix{}, \cosev{}, and \ci{} emission lines, as well as the underlying continuum emission of the quasar \qso. All the spectral lines and the continuum emission are spatially resolved at our resolution of $\sim 0\farcs4$. We show the intensity, velocity and velocity dispersion maps in Figure \ref{fig1}. The beam-matched \cii{} data is shown as well for comparison \citep{venemans20}.
We find a part with high velocity and high velocity dispersion northeast to the quasar on the CO and \cii{} maps. This position is coincident with the peak of the companion galaxy C1 discovered in \citet{venemans19}. The high velocity/velocity dispersion in that region is likely a result of interactions between the quasar and C1.
The velocity and velocity dispersion maps of all the spectral lines reveal some rotation and high velocity dispersion, which is consistent with results obtained for the super-high resolution \cii{} data \citep{venemans19}.
We measure the source sizes of the emission lines and the continuum emissions through the CASA UVMODELFIT task in the UV-plane. This leads to a comparable source size of (0.39 $\pm$ 0.04)$\times$(0.32 $\pm$ 0.04) with a position angle of PA=109\degree $\pm$ 75\degree{}  for the \cosix{} line, and (0.32 $\pm$ 0.03)$\times$(0.26 $\pm$ 0.03), PA=-68\degree $\pm$ 20\degree{}  for the \cosev{} line. 
The \cii{} line suggests a larger source size of (0.51 $\pm$ 0.02)$\times$(0.47 $\pm$ 0.02), PA=127\degree $\pm$ 28\degree.
 The deconvolved size for the continuum emission at 98.7 GHz is (0.30 $\pm$ 0.01)$\times$(0.28 $\pm$ 0.01), PA=9\degree $\pm$ 43\degree.

We measure the spectral line fluxes within an aperture of $ 0\farcs75$ with the residual scaling method presented in \citet{novak19}, to obtain emission of the quasar while avoiding possible contaminations from the close companion C1.  The resulting spectra are shown in Figure \ref{spectra3}. As the \ci{} emission line is not as strong as the CO lines, we assume the same line width for the CO and the \ci{} lines and fit the three lines, namely \cosix{},  \cosev{}, and \ci{}, simultaneously with three Gaussians. This leads to a line width of $250 \pm 11$ \kmps{} in FWHM, and line fluxes of 0.63 $\pm$ 0.04, 0.55 $\pm$ 0.04,  and 0.25 $\pm$ 0.03 \jykmps{} for the \cosix{},  \cosev{}, and \ci{} lines, respectively. The derived \cosix{} and  \cosev{} line fluxes are consistent with results obtained in ALMA Cycle 2 observations within the uncertainties \citep{venemans17a}.
We fit the spectrum of the \cii{} line with a Gaussian profile. This yields a line width of $268 \pm 11$ \kmps{} in FWHM, and a line flux of 5.25 $\pm$ 0.30 \jykmps. The derived line widths for the \cii{}, CO and \ci{} lines are consistent within the uncertainties.
The measured \cii{} flux is consistent with that obtained in \citet{venemans19} and \citet{venemans20}.
The measured line widths, fluxes, and luminosities are listed in Table \ref{tab:deluxesplit}. The continuum flux densities measured within the aperture are 0.27 $\pm$ 0.02  mJy and 5.20 $\pm$ 0.08 mJy at 98.7 and 258.1GHz, respectively.

\subsection{Companion galaxies}
\begin{figure*}
\centering
\gridline{\fig{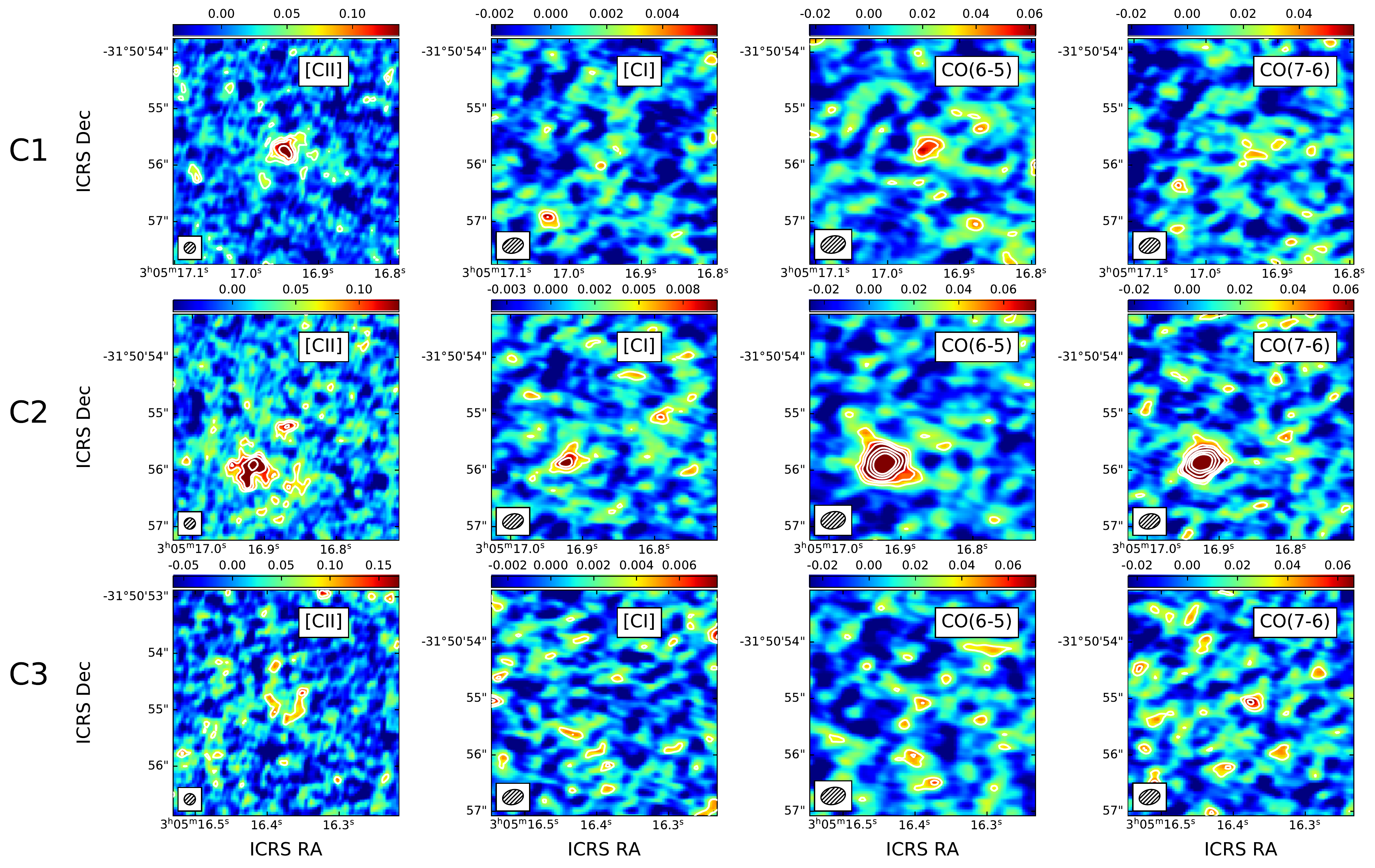}{1.0\textwidth}{(a)}}
\caption{\cii{},   \ci{}, \cosix{},  and \cosev{}  intensity maps (from left to right) centered on the companion galaxies C1,  C2 and C3 (from top to bottom). The beam is shown on the bottom left. 
Black contours on the C1 intensity maps are [--2, 2, 3, 4,  6] $\times \sigma$ ($ \sigma=0.024$ \jykmps{}),  [--2, 2] $\times \sigma$ ($ \sigma=0.012$ \jykmps{}), [--2, 2, 3] $\times \sigma$ ($ \sigma=0.012$ \jykmps{}),  and [--2, 2] $\times \sigma$ ($ \sigma=0.011$ \jykmps{}) from left to right. 
Contours on the C2 maps are [--2, 2, 3, 4] $\times \sigma$ ($ \sigma=0.031$ \jykmps{}),  [--2, 2] $\times \sigma$ ($ \sigma=0.012$ \jykmps{}),  [--2, 2] $\times \sigma$ ($ \sigma=0.013$ \jykmps{}), and [--2, 2] $\times \sigma$ ($ \sigma=0.010$ \jykmps{}) from left to right. 
As for C3,  black contours are [--2, 2, 3, 4] $\times \sigma$ ($ \sigma=0.041$ \jykmps{}),  [--2, 2] $\times \sigma$ ($ \sigma=0.012$ \jykmps{}),  [--2, 2, 3] $\times \sigma$ ($ \sigma=0.016$ \jykmps{}),  and [--2, 2, 3, 4] $\times \sigma$ ($ \sigma=0.017$ \jykmps{}) from left to right. \label{companionfig}}
\end{figure*}

\begin{figure*}
\centering
\gridline{\fig{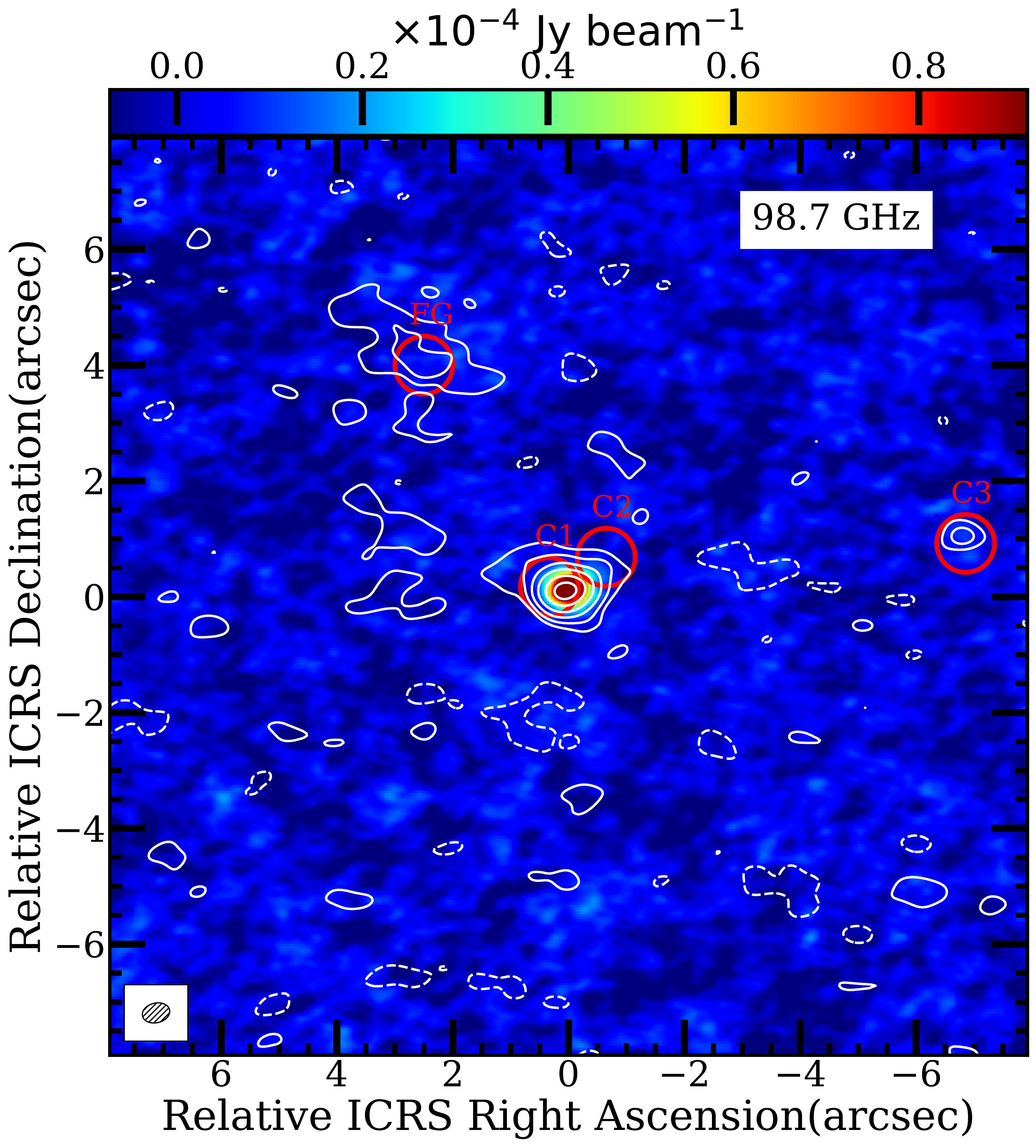}{0.8\textwidth}{(a)}}
\caption{Colors represent continuum at 98.7GHz. White contours: Beam-matched continuum at 258.1GHz with levels of [--2, 2, 4, 8, 16, 32, 55] $\times \sigma$ ($\sigma=23.8$ $\rm \mu\ Jy beam^{-1}$). Positions of the companion galaxies are shown as red circles.  `FG' represents the foreground source reported in Fig. 6 of \citealt{venemans19}. \label{continuum} 
}
\end{figure*}

Our re-analysis of the \cii{} data confirms the three companion sources originally reported in  \citep{venemans19}. In addition,  we detect the \cosix{} or/and \cosev{} lines in some of the companion galaxies (namely C1, C2, and C3).
The \cii{}, \ci{}, \cosix{}, and \cosev{}  intensity maps of C1, C2, and C3 are shown in Figure \ref{companionfig}. 
The spectral line fluxes for companion galaxies are measured in intensity maps through adopting line widths determined from super-high resolution \cii{} data \citep{venemans19}. 
We detect the \cii{} and \cosix{} lines in C1. We do not detect the \cosev{} line, possibly because of low S/N at that frequency.
Considering its close distance to the quasar, we measure the line flux within a 0\farcs3 radius aperture centered on the \cii{} peak. The resulting line fluxes are $0.06 \pm 0.01$ and $0.48 \pm 0.03$ \jykmps{} for \cosix{} and  \cii{}. The 3$\sigma$ upper limits for \cosev{} and \ci{} are both 0.04 \jykmps{}. 
C2 is only detected in the \cii{} line with a line flux of $0.27 \pm 0.07$ \jykmps{}, leaving 3$\sigma$ upper limits of 0.05, 0.04, and  0.04  \jykmps{} for \cosix{}, \cosev{}, and \ci{}. 
As for C3, the \cii{} intensity map suggests an extended gas structure, while the \cosix{} and \cosev{} emission are not spatially resolved. The extended \cii{} feature is also observed in the super-high resolution \cii{} data \citep{venemans19}. The \cosix{} and \cosev{} fluxes of C3 are $0.05 \pm 0.02$, and $0.07 \pm 0.02$ \jykmps. The \ci{} flux upper limit for C3  is 0.05 \jykmps.
The measured \cii{} fluxes for the three companion galaxies are consistent with those obtained in the high-spatial resolution observations \citep{venemans19}.
We show the spectra of \cii{} for the companion galaxies in Figure \ref{spectra3}.
As for the continuum detections, only C3 has been detected in the 258.1 GHz continuum, leaving a continuum flux density of 0.44 $\pm$ 0.08 mJy. Through combining observations in different ALMA cycles, \citet{venemans20} detect the 258.1GHz continuum emission in all the three companion galaxies around \qso{}.  The non-detection of the C2 continuum at 258.1GHz thus suggests that the 3$\sigma$ limit of the emission surface brightness is 16.0$\rm \mu Jy\ kpc^{-2}$.  The 3$\sigma$ upper limits for the surface brightness for C2 and C3 at 98.7GHz are both 1.5$\rm \mu Jy\ kpc^{-2}$.
In Figure \ref{continuum}, we show the continuum maps at 98.7 and 258.1GHz. The peak positions of the \cii{} line and the 258.1GHz continuum for C3 are consistent.  Details of the quasar and the companion galaxy measurements are listed in Table \ref{tab:deluxesplit}.

\begin{table*}[t]
\scriptsize
\centering
\caption{Observational results and derived properties of the quasar \qso{} and its companion galaxies}
\begin{tabular}{lcccc}
\hline 
\hline
{} & \qso & C1 & C2 &C3 \\
\hline
R.A.  & 03$^\mathrm{h}$05$^\mathrm{m}$16$^\mathrm{s}\!\!$.92
& 03$^\mathrm{h}$05$^\mathrm{m}$16$^\mathrm{s}\!\!$.95
& 03$^\mathrm{h}$05$^\mathrm{m}$16$^\mathrm{s}\!\!$.87 
& 03$^\mathrm{h}$05$^\mathrm{m}$16$^\mathrm{s}\!\!$.38
 \\
Decl. & --31$^\circ$50\arcmin55\farcs92(6) 
& --31$^\circ$50\arcmin55\farcs74(13) 
& --31$^\circ$50\arcmin55\farcs26(23)
& --31$^\circ$50\arcmin54\farcs96(31) 
 \\
$z_\mathrm{[CII]} $  & 6.61391$\pm$0.00015
& 6.6231$\pm$0.0003
&6.6104$\pm$0.0004 
&6.6066$\pm$0.0006 \\
$\rm FWHM_\mathrm{CO}$ (\kmps) &  250 $\pm$ 11 \\
\cosix{} flux (\jykmps) & 0.63 $\pm$ 0.04 & 0.06 $\pm$ 0.01  & $< 0.05$& 0.05 $\pm$ 0.02\\
$L_\mathrm{CO(6-5)}$ ($\times 10^{8}$ \lsun) &  2.49 $\pm$ 0.16&0.25 $\pm$ 0.04&$<$0.18&0.20 $\pm$ 0.08\\
Deconvolved size ($\mathrm{CO(6-5)}$)&  (0.39 $\pm$ 0.04)$\times$(0.32 $\pm$ 0.04), PA=109\degree $\pm$ 75\degree    \\
\cosev{} flux (\jykmps) &0.55 $\pm$ 0.04 &$< 0.04$&$< 0.04$&0.07 $\pm$ 0.02 \\
$L_\mathrm{CO(7-6)}$ ($\times 10^{8}$ \lsun) &2.53 $\pm$ 0.18&$<$0.19& $<$0.20& 0.31 $\pm$ 0.08\\
Deconvolved size ($\mathrm{CO(7-6)}$)&  (0.32 $\pm$ 0.03)$\times$(0.26 $\pm$ 0.03), PA=-68\degree $\pm$ 20\degree \\
\ci{} flux (\jykmps) & 0.25 $\pm$ 0.03 &$< 0.04$ & $< 0.04$&$< 0.05$\\
$L_\mathrm{[CI](2-1)}$ ($\times 10^{8}$ \lsun) &1.15 $\pm$ 0.14&$<$0.19&$<$0.20&$<$0.22  \\
$S_\mathrm{258.1 \rm GHz}$ (mJy) &5.20 $\pm$ 0.08  && &0.44 $\pm$ 0.08\\
$S_\mathrm{98.7 \rm GHz}$ (mJy) &0.27 $\pm$ 0.02 & \\
Deconvolved size ($\mathrm{98.7 \rm GHz}$)&  (0.30 $\pm$ 0.01)$\times$(0.28 $\pm$ 0.01), PA=9\degree $\pm$ 43\degree\\
$\rm FWHM_\mathrm{[CII]}$ (\jykmps)&268 $\pm$ 11&\\
\cii{} flux (\jykmps) & 5.25 $\pm$ 0.30 &0.48 $\pm$ 0.03&0.27  $\pm$ 0.07 &0.87  $\pm$ 0.31  \\
$L_\mathrm{[CII]}$ ($\times 10^{8}$ \lsun)&56.94 $\pm$ 3.25& 5.16 $\pm$ 0.33&2.93 $\pm$ 0.76& 9.41 $\pm$ 3.36 \\
Deconvolved size $^e$ ($\mathrm{[CII]}$)&(0.51 $\pm$ 0.02)$\times$(0.47 $\pm$ 0.02), PA=127\degree $\pm$ 28\degree\\
\hline
$M_{\mathrm{H}_2,\mathrm{CO}}$ ($\times 10^{10}$ \msun) &2.2--4.0&0.21--0.37 &$<$0.36&0.24--0.44\\
$M_{\mathrm{H}_2,\mathrm{[CI]}}$ ($\times 10^{10}$ \msun)&3.0 $\pm$ 1.3&$<$0.48& $<$0.44&$<$0.60\\
$\Sigma_{\mathrm{H}_2,\mathrm{CO}}$ ($\rm M_{\sun} \ pc^{-2}$) & 1.4--2.6$\times 10^{4}$ &$> 213 $&&$> 244 $\\
$\Sigma_{\mathrm{H}_2,\mathrm{[CI]}}$ ($\rm M_{\sun} \ pc^{-2}$) & $(2.0 \pm 0.9) \times 10^{4}$&& \\
\hline
\end{tabular} 
\tablecomments{The source positions are determined from the peak pixels in the \cii{} intensity maps,  the astrometric uncertainties are given in mas in parentheses.  The \cii{} redshifts are adopted from \citet{venemans19}.  We assume the same line widths for the \cosix{}, \cosev{}, and \ci{} lines and fit the three lines together during the line fluxes measurements for the quasar, and the fitting result for the line width is given in $\rm FWHM_\mathrm{CO}$.
The fluxes for the quasar are measured within an aperture with a radius of 0\farcs75. To avoid contamination from the quasar, the flux of C1 is measured from  a 0\farcs3 radius aperture. We use 0\farcs6 radius aperture to obtain the \cii{} fluxes of the C3  emission line and continuum. To calculate the gas surface densities of the companion galaxies, we assume the emission to be spatially unresolved and apply the beam sizes as  upper limits for the source sizes.
}
\label{tab:deluxesplit}
\end{table*}
\section{Discussions}
\subsection{Spatial Distribution and resolved ratios of different gas tracers}
\begin{figure*}
\includegraphics[width=0.5\textwidth]{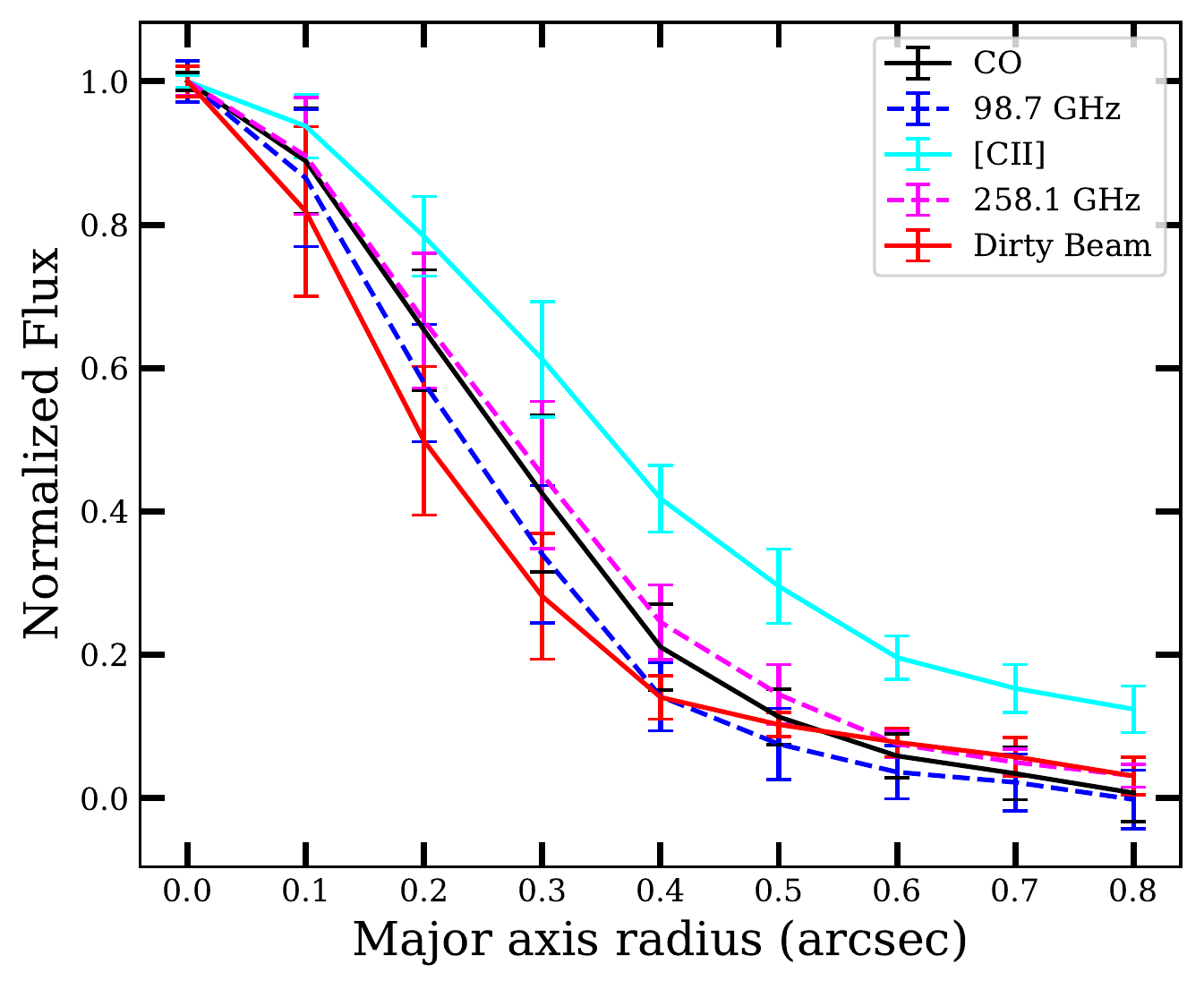}
\includegraphics[width=0.5\textwidth]{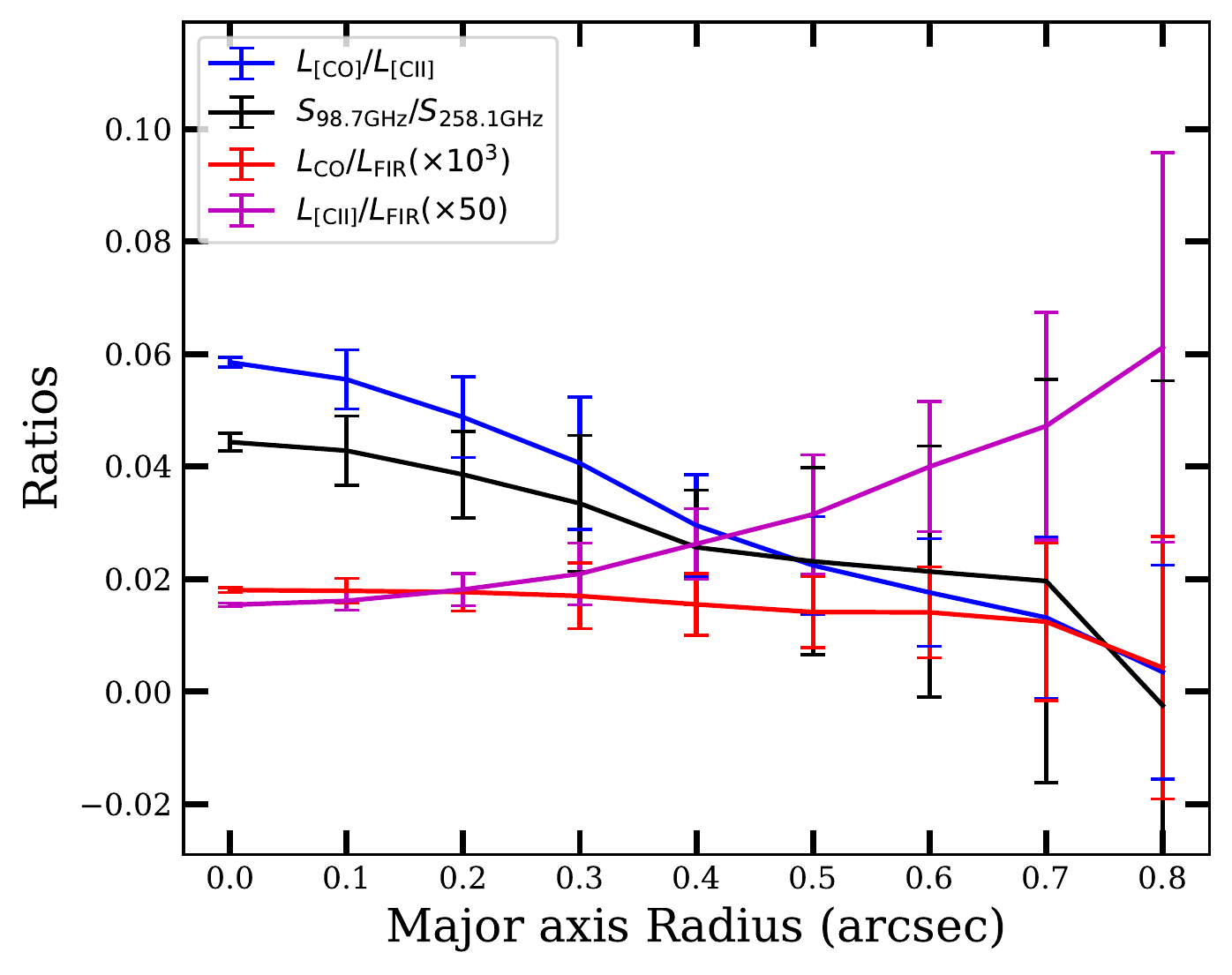}
\caption{Left: Radial profiles of different gas and dust continuum tracers. The line fluxes are normalized to the peak pixel fluxes. 
Right: Similar as the left panel, but showing the radio profiles of different line ratios.
\label{fig:radial}}
\end{figure*}

We study the dependence of the spectral line and the continuum intensities with distances to the quasar, to explore the spatial distribution of the gas in different phases and the dust.  To obtain the radial profiles, we divide the intensity maps into a series of concentric rings with a width of $0\farcs1$, with the center fixed to the peak flux pixel, and the major, the minor axes and the position angle fixed to the parameters adopted from the \cii{} source size.  To increase the S/N ratio of the CO measurements, we take the average of the \cosix{} and \cosev{} datacubes to form a mean CO intensity map.  The resulting radial profiles are shown in Figure \ref{fig:radial}.  
All the lines and continuum show intensities that exceed the radial profile of the dirty beam (0\farcs4) at large radii, implying that the sources are extended.
The radial profile of CO follows that of the continuum within the uncertainties, indicating that they are possibly originated from the same gas component.
The \cii{} radial profile reduces to half of the peak intensity at larger radius compared to that of the CO and dust continuum, which implies that the \cii{} emission is more extended than the CO and the continuum. 
This result is consistent with previous findings that the \cii{} line in $z\gtrsim 6$ quasars have larger source sizes than that of the mid to high-J (J $\gtrsim 5$) CO and dust continuum (e.g., \citealt{shao17}; \citealt{li20a}; \citealt{venemans20}).  As for  low-J (J $\lesssim 3$) CO lines,   \citet{shao19} find similar source sizes as that of \cii{} in a sample of three $z\gtrsim 6$ quasars, while both of which are larger than the sizes of dust continuum.

In Figure \ref{fig:radial}, we also show the radial profiles of the CO/\cii{}, CO/FIR, \cii{}/FIR ratios as well as the continuum flux density ratio ($S_{98.7 \rm GHz}/S_{258.1 \rm GHz}$). 
Similar to findings in high resolution \cii{} observations of other $z\gtrsim 6$ quasars, the \cii{}/FIR ratio  of \qso{} exhibits a deficit in the center and an increasing trend with increasing distance to the center. The  CO/FIR ratio is almost flat with increasing radius.  A decreasing trend of the CO/\cii{} ratio with increasing radius is consistent with an extended spatial distribution of the \cii{} line relative to CO assuming the spatial distribution of these two lines are Gaussian. Interestingly, we find a decreasing trend of the $S_{98.7 \rm GHz}/S_{258.1 \rm GHz}$ ratio with increasing radius. 

The dust emission ($S_{\nu}$) is described through 
\begin{align}
S_{\nu} \propto (1-e^{-\tau_{\nu}})(B_{\nu}(T_{dust})-B_{\nu}(T_{CMB})),
\end{align}
where $B_{\nu}(T_{dust})$ and $B_{\nu}(T_{CMB})$ are the Planck function at dust temperature and Cosmic Microwave Background temperature,   respectively.  $\tau_{\nu} $ is optical depth which can be further expressed as 
\begin{align}
\tau_{\nu} = \kappa_{\nu} \Sigma_{dust}. 
\end{align}
$ \kappa_{\nu}$ is dust opacity,  which depends on frequency through
\begin{align}
 \kappa_{\nu} = \kappa_{0}(\frac{\nu}{\nu_{0}})^{\beta},
\end{align}
and $\Sigma_{dust}$ is dust mass surface density, which is independent on frequency.  
We consider two simplifications.  In the first case we assume $\tau_{\nu} $ remain constant with radius,  but we enable $T_{dust}$ as a function of radius ($r$).  The derivatives of the dust continuum ratio ($S_{98.7 \rm GHz}/S_{258.1 \rm GHz}$) relative to $r$ is :
\begin{align}
\frac{d(S_{98.7 \rm GHz}/S_{258.1 \rm GHz})}{d(r)} \propto - \frac{d(T_{dust}(r))}{d(r)},
\end{align}
suggesting an opposite monotonicity of $S_{98.7 \rm GHz}/S_{258.1 \rm GHz}$ relative to $T_{dust}(r)$. 
In the second simplification, we assume a constant $T_{dust}$ throughout the source, while consider $\tau_{\nu}$ as a function of $r$ (the $r$ dependence of $\tau_{\nu}(r) $ reduces to $\Sigma_{dust}(r)$ through $\tau_{\nu}(r) = \kappa_{0}(\frac{\nu}{\nu_{0}})^{\beta}\times \Sigma_{dust}(r)$).
The derivative of $S_{98.7 \rm GHz}/S_{258.1 \rm GHz}$ relative to $r$ is:
\begin{align}
\frac{d(S_{98.7 \rm GHz}/S_{258.1 \rm GHz})}{d(r)} \propto \frac{d(\tau_{\nu}(r))}{d(r)} \propto \frac{d(\Sigma_{dust}(r))}{d(r)}, 
\end{align}
indicating the same monotonicity between $S_{98.7 \rm GHz}/S_{258.1 \rm GHz}$ and  $\tau_{\nu}(r)$ (or $\Sigma_{dust}(r)$).
Accordingly, the decreasing $S_{98.7 \rm GHz}/S_{258.1 \rm GHz}$ ratio with increasing radius can be explained by (1) an increase of $T_{dust}(r)$ with $r$ assuming a uniform $\tau_{\nu} $ or (2) a decrease of $\tau_{\nu}(r)$ or $\Sigma_{dust}(r)$ with $r$ for constant $T_{dust}$ across $r$.  
An increasing temperature with increasing radius seems inconsistent with the scenario of more AGN dust heating toward the center. However, we cannot rule out that the interactions between the quasar and C1 that drive the observed increasing temperature with increasing distances to the center. 
The second explanation is more likely to be the case,  given that expected decreasing  $\Sigma_{dust}$ (or $\tau_{\nu}(r)$) with radius.

\subsection{Gas properties}
We estimate the molecular gas mass of the quasar based on the CO and \ci{} lines. 
Assuming the \ci{} emission to be optically thin, the neutral carbon mass can be estimated through 
\begin{align}
\frac{M_{\rm C}}{M_{\sun}} = 4.566\times 10^{-4} \frac{Q(T_{\rm ex})}{5} e^{62.5/T_{\rm ex}} \frac{L_{\rm [CII]}^{'}}{\rm K\ km\ s^{-1}\ pc^{-2}},
\end{align}
where $Q({T_{\rm ex}}) = 1+ 3e^{-23.6/T_{\rm ex}} + 5e^{-62.5/T_{\rm ex}}$ is the partition function and $T_{\rm ex}$ is the excitation temperature (\citealt{ weiss03,weiss05}; \citealt{venemans17a}). Adopting $T_{\rm ex}=30$ K from \citealt{venemans17a}, we estimate an neutral carbon mass of $\rm M_{[CI]}= (1.5\pm 0.2) \times 10^7 $ \msun{} based on the \ci{} line. The neutral carbon abundance relative to molecular hydrogen is $X_{[CI]} = \rm M_{[CI]}/(6\ M_{H_{2}})$. Utilizing the neutral carbon abundance measured in $z=2-3$ infrared bright galaxies from \citet{walter11} of $X_{[CI]} = (8.4 \pm 3.5) \times 10^{-5}$, we estimate a molecular gas mass of $(3.0 \pm 1.3) \times 10^{10} $ \msun. 
We derive a molecular gas mass surface density of $(2.0 \pm 0.9) \times 10^{4} $ $\rm M_{\sun} \ pc^{-2}$.
We also estimate the molecular gas masses for the three companion galaxies based on their \ci{} upper limits. The resulting 3$\sigma$ upper limits for the molecular gas masses of C1, C2 and C3 are $<$4.8$\times 10^9$\msun{}, $<$4.4$\times 10^9$\msun{} and $<$6.0$\times 10^9$\msun{}, respectively.

Low-$J$ CO transitions trace the cold molecular gas, and thus can be used as molecular gas mass indicators through
\begin{align}
M_{H_{2}}=\alpha_{co} \times L^{'}_{co}.
\end{align}
Where $\alpha_{co}$ is the molecular gas conversion factor, and $L^{'}_{co}$ is the CO luminosity in the unit of $\rm K\ km\ s^{-1}\ pc^{2}$. $L^{'}_{co}$ is calculated from the CO flux $S_{co}\delta v $ through
\begin{align}
L^{'}_{co}=3.25\times 10^{7}S_{co}\delta v \frac{D_{L}^{2}}{(1+z)^{3}\nu_{obs}^{2}}.
\end{align}
For \qso{}, we only detect the \cosix{} and \cosev{} lines. To derive molecular gas mass, we utilize the following two methods to estimate the \coone{} flux from the \cosix{} line. (1) We employ the \cosix{}/\coone{} ratio of the CO spectral line energy distribution model prediction for J2310+1855 \citep{li20a}, this leads to an estimated molecular gas mas of $(2.70 \pm 0.17) \times 10^{10} $ \msun. (2) We use the observed \cosix{}/\cotwo{} flux ratio in the range of 5.7--10.3 for $z\gtrsim 6$ quasars in \citet{shao19}. 
Using the approximation that $L^{'}_{CO(2-1)}\approx L^{'}_{CO(1-0)}$ from \citet{carilli13}, and adopting a conversion factor for local (Ultra) Luminous Infrared Galaxies ((U)LIRGs) of $\alpha_{co}=0.8\  \rm M_{\odot} (K\ km\ s^{-1}\ pc^{2})^{-1}$\citep{downes98}, we estimate the molecular gas mass of \qso{} from the CO lines to be 2.2--4.0 $\times 10^{10}$\msun. The derived molecular gas mass surface density is 1.4--2.6$\times 10^{4} $ $\rm M_{\sun} \ pc^{-2}$.
As for the companion galaxies, the molecular gas masses based on CO are 2.1--3.7 $\times 10^9$\msun{}, $<$3.6$\times 10^9$\msun{} and 2.4--4.4 $\times 10^9$\msun{} for  C1, C2 and C3, respectively.  Assuming the CO lines of companion galaxies are spatially unresolved, we thus estimate the 3$\sigma$ lower limit of the gas mass surface density to be $> 213 $ $\rm M_{\sun} \ pc^{-2}$ and $> 244  $ $\rm M_{\sun} \ pc^{-2}$ for C1 and C3, respectively.

The gas masses based on CO and \ci{} are within the uncertainties. 
In Figure \ref{fig:def}, we show the relation between the star formation rate density and molecular gas surface density for \qso{} and the companion galaxies.
The derived molecular gas mass surface densities of the quasar from both CO and \ci{} are consistent and comparable to the maximum values found in local starburst galaxies \citep{kennicutt21}. The gas surface density 3$\sigma$ upper limits for  C3 is consistent with the lowest value found for the average of local starburst galaxies, and higher than that of local spiral galaxies (\citealt{kennicutt19}; \citealt{kennicutt21}).  Similar as that found for other high-$z$ quasars and galaxies (e.g., J0100, BRI 1202 QSO, and BRI 1202 SMG),  \qso{} and the companion galaxy C3 reside on the local star-formation law.
The molecular gas masses for all the companion galaxies are an order of magnitude lower than that in the quasar.


\subsection{The ISM properties of the quasar and its close companions}
The ISM lines and continuum emission at (sub)millimeter provide rich information on the ISM properties.  For example, when the illumination radiation field is dominated by X-rays the ISM emission in X-ray dominated regions (XDRs) tend to have lower \cii{}/\ci{} ratio compared to a radiation field dominated by UV photons (the ISM components illuminated by UV photons are generally referred to as photo-dissociation regions, PDRs).  In local AGNs and (U)LIRGs, extensive observations of the FIR fine structure lines suggest a deficit of line-to-FIR ratio with increasing FIR luminosity, namely the ``FIR line deficit".   High CO-to-FIR ratios are generally expected when the ISM is heated by e.g., X-rays or shocks besides UV photons from young massive stars \citep{uzgil16}.
In this work, we detect the CO, \cii{}, \ci{} and continuum emissions in the quasar \qso{} and its three close  companions. 
Direct comparisons between the quasar and companion galaxies thus enable us to explore the possible impacts of the central accreting supermassive black hole on the ISM.

\begin{figure*}
\includegraphics[width=0.5\textwidth]{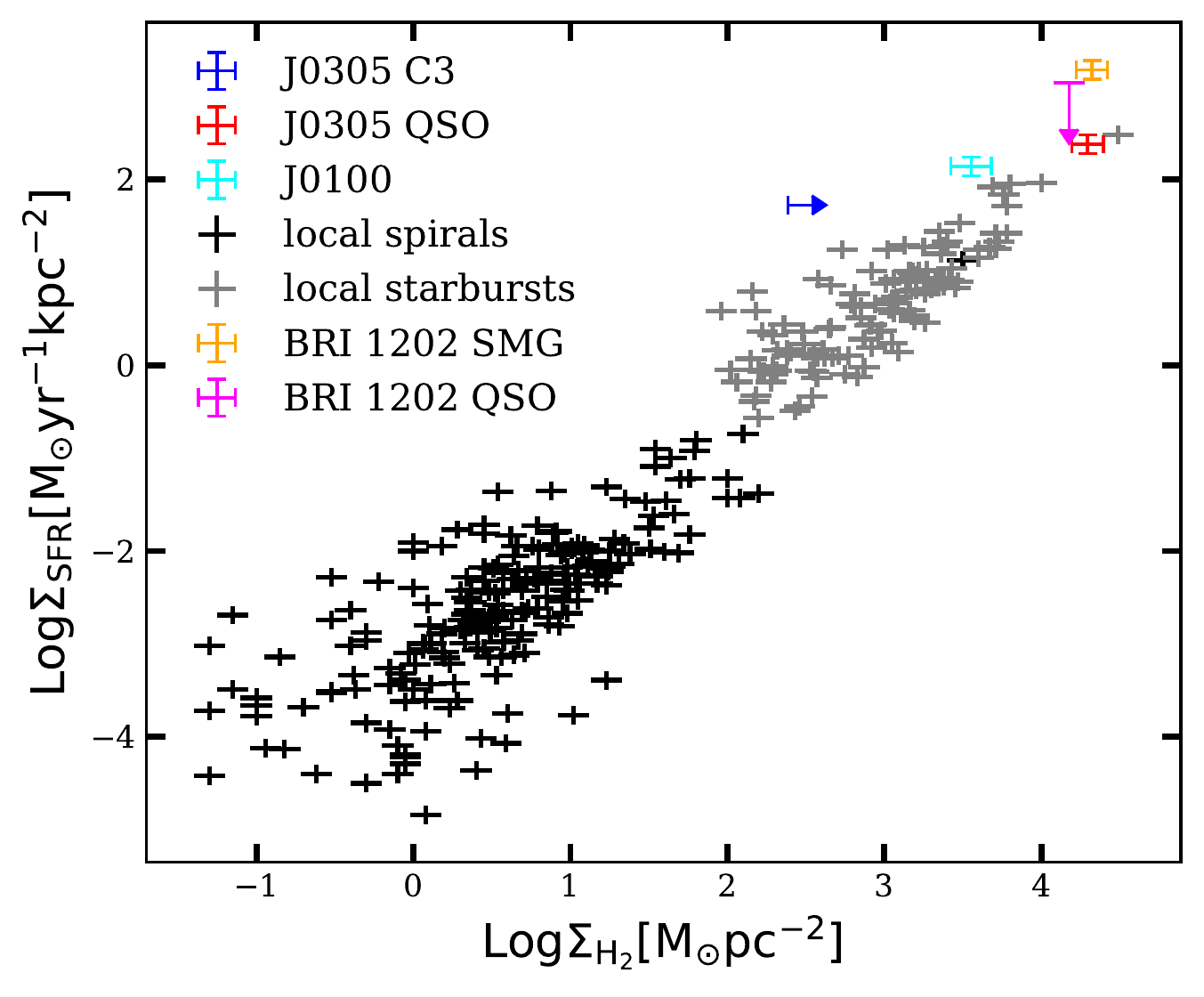}
\includegraphics[width=0.5\textwidth]{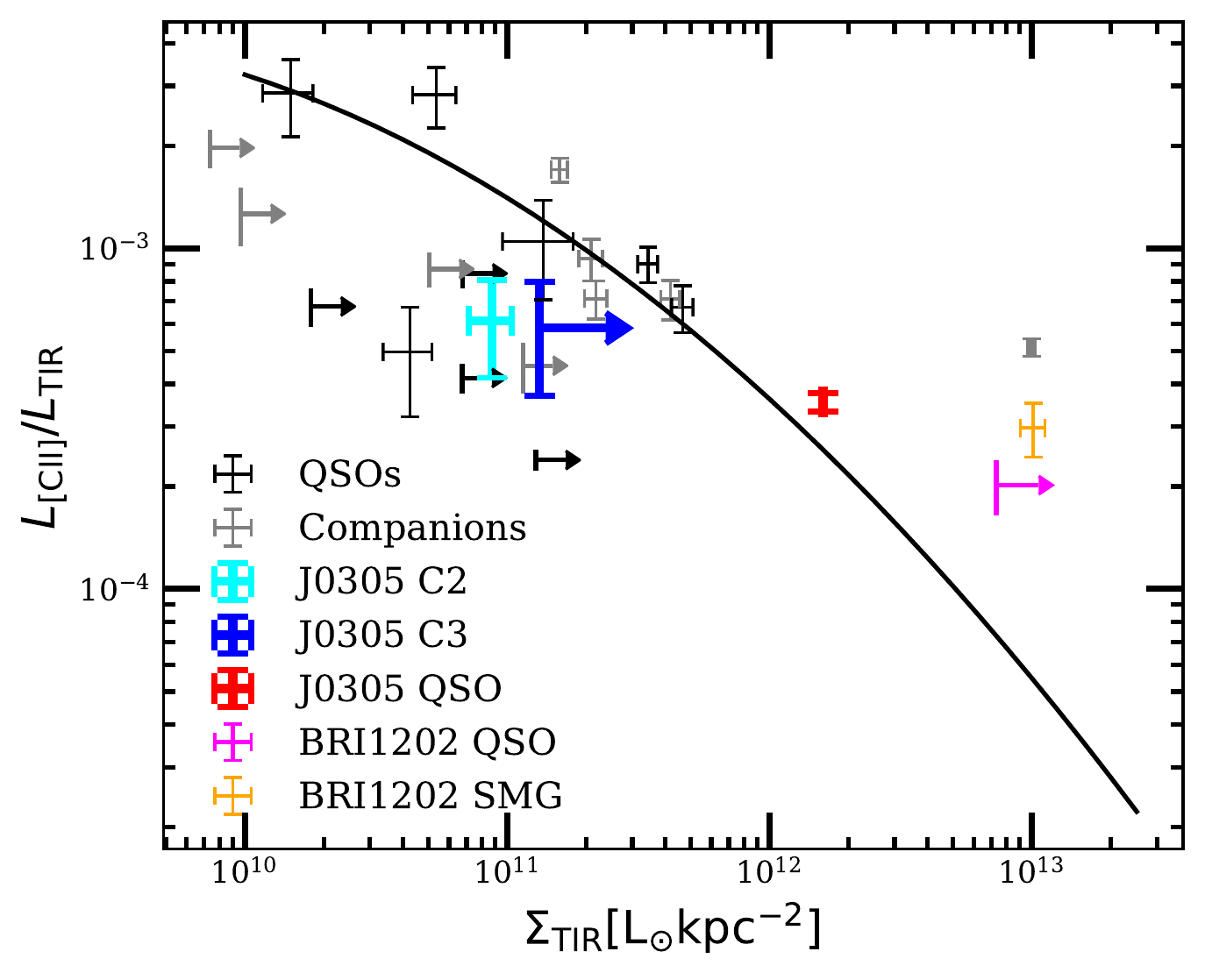}
\caption{Left:  SFR surface density $vs$ molecular gas surface density for \qso{} and its companion galaxy C3 (the molecular gas surface densities of C1 and C2 are unable to determine).  
Comparison samples include local spiral  (``local spirals") and starburst (``local starbursts") galaxies  from \citet{kennicutt19} and \citet{kennicutt21},  the z=4.69 quasar BRI 1202 (``BRI 1202 QSO") and its companion submillimeter galaxy (``BRI 1202  SMG") from \citet{wagg12} and \citet{jones16},  and the z=6.33 quasar J0100+2802(``J0100") from \citet{wang19b}. The SFRs of \qso{} and C3 are calculated using the same formula as that presented in  \citet{kennicutt21}.
Right: \cii{} deficit diagram for the quasar \qso{} and its companion galaxy C2 and C3 (C1 is not included because of difficulties in constraining the source size).   Assuming the CO lines of C3 to be spatially unresolved, we use the beam size of \cosix{} as an upper limit for the source size in the TIR surface density estimation of C3.
 ``QSOs" and ``companions" refer to z$\sim$6 quasar and companion galaxy samples collected from 
 \citet{neelman19} and \citet{decarli17}. 
The black solid line represents the \cii{} deficit relation observed in local LIRGs from 
\citet{diaz17}. For the objects in this work and in the comparison samples, the total infrared (TIR) luminosities are estimated from the continuum flux density close to the \cii{}  line frequency through assuming a modified black body model with temperature of 47K and emissivity of 1.6. 
\label{fig:def}}
\end{figure*}

Adopting a FIR dust continuum modified black body model with parameters of $T_{dust}=47\ K$ and $\beta=1.6$ \citep{venemans19},  we calculate the FIR luminosities based on the \cii{} continuum flux densities for the quasar and the companion galaxies. We obtain \cosix{},  \cosev{}, \ci{}, and \cii{}-to-FIR ratios of $2.3\times10^{-5}$, $2.3\times10^{-5}$, $1.4\times10^{-5}$, and $5.0\times10^{-4}$ for \qso{}.   
We adopt dust continuum flux densities in \citet{venemans20} for FIR luminosity estimations for C1, C2, and C3.
We estimate the \cosix{}, \cosev{},\ci{}, and \cii{} to FIR luminosity ratios for C1 to be $(1.8\pm 0.5) \times10^{-5}$,  $< 1.4 \times10^{-5}$, $< 1.4 \times10^{-5}$, and  $(4.1 \pm 0.9)  \times10^{-3}$.
The  \cosix{}, \cosev{},\ci{}, and \cii{} to FIR luminosity ratios of C2 are $< 5.3 \times10^{-5}$, $< 5.9 \times10^{-5}$, $< 5.9 \times10^{-5}$, and $(8.6 \pm 0.3)  \times10^{-4}$.
As for C3, the ratios between \cosix{}, \cosev{}, \ci{}, \cii{} and FIR luminosity are $(1.8\pm 0.7) \times10^{-5}$,  $(2.7\pm 0.7) \times10^{-5}$,  $(1.8\pm 0.5) \times10^{-5}$, $< 1.9 \times10^{-5}$,  and $(8.3\pm 0.3) \times10^{-4}$.
We also calculate the total infrared (TIR; 8$\sim$1000 $\mu m$) surface densities and \cii{}/TIR ratios of the quasar, and its companion galaxy C2 and C3 for a direct comparison with other samples.  
In Figure \ref{fig:def}, we show the \cii{}/TIR $vs$ $\Sigma_{\rm TIR}$.  Similar as those found for other $z \gtrsim 6$ quasars and companion galaxies, \qso{} and its companion galaxies follow the spatially resolved \cii{} deficit trend of local LIRGs (\citealt{wagg12}; \citealt{jones16}; \citealt{decarli17};  \citealt{diaz17}; \citealt{neelman19}).

In addition, we calculate the \cii{}/\ci{} ratios of the quasar \qso{} and its close companions as diagnostics between PDRs and XDRs. 
The quasar \qso{} exhibits a \cii{}/\ci{} ratio of 49.5, which is consistent with PDRs \citep{meijerink05,meijerink07}. The lower limits of the \cii{}/\ci{} ratios for C1, C2 and C3 are 36.9, 19.5 and 49.6 respectively, which is also within the ranges for the PDR model prediction.  Despite the presence of luminous AGN, the \cii{}/\ci{} ratio of \qso{} excludes the XDR dominance in the ISM excitation of the quasar.
\qso{} reveals a \cii{}-to-CO ratio of $\sim$ 10. Comparable ratios are found for the companion galaxies. To summarize, we find no significant differences in the line-to-line and line-to-FIR ratios between \qso{} and its companion galaxies.


\section{Summary}
In this work, we present an analysis of the \cosix{}, \cosev{}, \ci{} lines as well as the dust continuum emission in the quasar \qso{} at 0\farcs4 resolution. Some of these lines are also detected in the companion galaxies within 40 kpc from the quasar. We summarize the main results below.

$\bullet$ We detect CO emission in two of the three companion galaxies.  Their respective CO fluxes are an order of magnitude fainter than those observed in the quasar. We derive molecular gas masses and molecular gas mass surface densities for the quasar and companion galaxies from both CO and \ci{} luminosities. 
The gas mass in the quasar is an order of magnitude higher than those found for companion galaxies. 
The molecular gas mass surface density of the quasar is at the high end of what is found in local starburst galaxies.  The upper limits for the gas mass surface densities of the companion galaxies are comparable to the lowest values found in local starburst galaxies and higher than local spiral galaxies.

$\bullet$ We compare the radial profiles of the CO, \cii{} and the dust continuum emission of the quasar. 
The \cii{} profile is above both the CO and the dust continuum, suggesting a more extended spatial distribution of \cii{} relative to the CO and the dust continuum. CO and dust continuum have similar radial profiles, implying similar gas component as traced by the CO and the dust.
In addition, we calculate the CO/\cii{}, \cii{}-to-FIR, CO-to-FIR, and dust continuum $S_{98.7 \rm GHz}/S_{258.1 \rm GHz}$ ratio profiles. The CO/\cii{} ratio decreases with increasing radius, confirming the more extended spatial distribution of CO compared to \cii{}. The decreasing \cii{}-to-FIR ratio with increasing distance to the center is consistent with high-resolution \cii{} observations of other $z\gtrsim 6$ quasars. 
The CO-to-FIR ratio on the other hand is almost flat with radius. We find a decreasing $S_{98.7 \rm GHz}/S_{258.1 \rm GHz}$ ratio with increasing radius, which is possibly due to a decrease of dust optical depth with increasing radius.

$\bullet$ We compare the ISM properties in the quasar and companion galaxies through the \cii{}/\ci{}, CO/\cii{}, \cii{}-to-FIR, and CO-to-FIR ratios. No significant differences are found between the quasar and the companion galaxies. Future high-J (J$\geq$10) CO observations will be critical in discriminating the differences of ISM properties between the quasar and the companions.

\section{Acknowledgements}
We thank Mladen Novak and Melanie Kaasinen for help with the data reduction.  We thank the referee for constructive comments that helped improve our manuscript.
This work was supported by the National Science Foundation of China (NSFC, 11721303, 11991052) and the National Key R\&D Program of China (2016YFA0400703). R.W. acknowledges supports from the NSFC grants No. 11533001 and the Thousand Youth Talents Program of China.  
B.P.V. and F.W. acknowledge funding through ERC Advanced Grant 740246 (Cosmic Gas).
This paper makes use of the following ALMA data: ADS/JAO.ALMA$\#$2017.1.01532.S, ADS/JAO.ALMA$\#$2015.1.00399.S. 
ALMA is a partnership of ESO (representing its member states), NSF (USA) and NINS (Japan), together with NRC (Canada), MOST and ASIAA (Taiwan), and KASI (Republic of Korea), in cooperation with the Republic of Chile. The Joint ALMA Observatory is operated by ESO, AUI/NRAO and NAOJ.

\end{document}